\documentclass{webofc}
\usepackage[varg]{txfonts}   % Web of Conferences font
\begin{document}
\title{Highlights from STAR heavy ion program}
%\subtitle{QFTHEP 2017}

\author{\firstname{Vitalii} \lastname{Okorokov}\inst{1}\fnsep\thanks{\email{VAOkorokov@mephi.ru;    Okorokov@bnl.gov}} (on behalf of the STAR Collaboration)
}

\institute{National Research Nuclear University MEPhI (Moscow Engineering Physics Institute), \\ Kashirskoe avenue 31, 115409 Moscow, Russian Federation}

\abstract{
Recent experimental results obtained in STAR experiment at the Relativistic heavy-ion collider (RHIC) with ion beams will be discussed. Investigations of different nuclear collisions in some recent years focus on two main tasks, namely, detail study of quark-gluon matter properties and exploration of the quantum chromodynamics (QCD) phase diagram. Results at top RHIC energy show clearly the collective behavior of heavy quarks in nucleus-nucleus interactions. Jet and heavy hadron measurements lead to new constraints for energy loss models for various flavors. Heavy-ion collisions are unique tool for the study of topological properties of theory as well as the magneto-hydrodynamics of strongly interacting matter. Experimental results obtained for discrete QCD symmetries at finite temperatures confirm indirectly the topologically non-trivial structure of QCD vacuum. Finite global vorticity observed in non-central Au+Au collisions can be considered as important signature for presence of various chiral effects in sQGP. Most results obtained during stage I of the RHIC beam energy scan (BES) program show smooth behavior vs initial energy. However certain results suggest the transition in the domain of dominance of hadronic degrees of freedom at center-of-mass energies between 10--20 GeV. The stage II of the BES at RHIC will occur in 2019--2020 and will explore with precision measurements in the domain of the QCD phase diagram with high baryon densities. Future developments and more precise studies of features of QCD phase diagram in the framework of stage II of RHIC BES will be briefly discussed.
}
\maketitle
\section{Introduction} \label{sec-1}

\begin{figure}
\centering
\sidecaption
\includegraphics[width=8cm,clip]{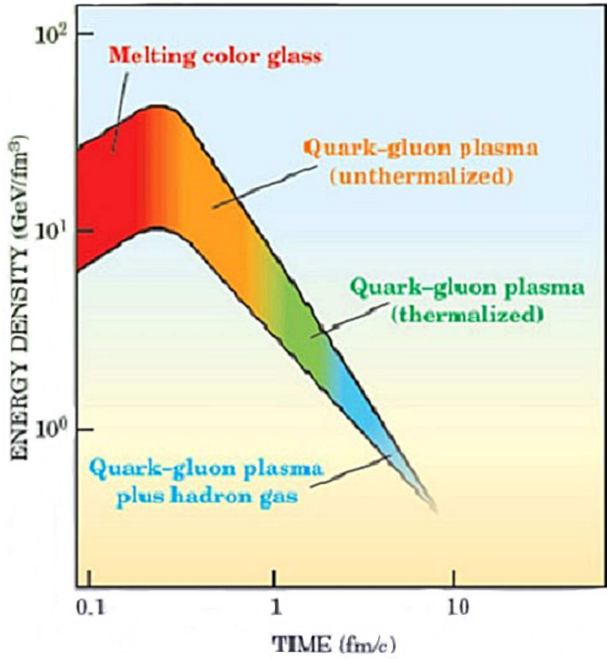}
\caption{Cartoon for evolution of strongly interacting matter in nuclear collisions. Picture is taken from \cite{Ludlam-PT-56-48-2003}.}\label{fig-1-1-1}
\end{figure}

Numerous experimental results have suggested that nucleus-nucleus collisions produce hot matter consisting of deconfined quark and gluons. This new state of matter behaves like (quasi)ideal quark-gluon liquid and the matter is called strongly-coupled quark-gluon plasma (sQGP). The sQGP is observed in wide energy domain, at least for $\sqrt{s_{NN}} \gtrsim 39$ GeV, where $\sqrt{s_{NN}}$ is the collision energy per nucleon-nucleon pair. The study of the
transition from hadronic matter to the quark-gluon state as well as properties of sQGP is one of the most important and debated problems of modern physics of strong interactions. The final state created in nucleus-nucleus collision evolves through several stages \cite{Ludlam-PT-56-48-2003}.
The intensive investigations within relativistic nuclear physics during last decades provided the some general framework for space-time evolution of nuclear collisions called also "Standard Model" for (heavy) ion collisions (Fig. \ref{fig-1-1-1}). According to this conception nucleus-nucleus collisions at relativistic energies can be considered as the collision of two sheets of strongly interacting matter appeared due to Lorentz contraction \cite{Kovner-PRD-52-6231-1995}. The initial state matter characterized by high energy density and high degree of coherence is the coherent limit of QCD at high energies. Thus this gluonic matter is called color glass condensate (CGC) \cite{Iancu-arXiv-hep-ph-0303204-2003}. At high energies, these sheets pass through one another, the fields become singular and CGC melts with production of the glasma \cite{Kovner-PRD-52-6231-1995,Krasnitz-PRL-86-1717-2001} which eventually materializes as quarks and gluons. It involves many strongly interacting and highly coherent color electric and color magnetic longitudinal flux lines at relatively low baryon densities. It is important to emphasize that because there are both longitudinal color electric and longitudinal color magnetic fields, there is a non-zero density of topological charge, $\vec{E} \cdot \vec{B}$ \cite{McLerran-arXiv-0911.2987}. A non-zero average of such a quantity would signal the breaking of fundamental QCD symmetries ($\mathcal{P}$ and $\mathcal{CP}$). The one of the possible mechanisms for violation of these symmetries are the dynamical topological sphaleron transitions which may result in observable metastable $\mathcal{CP}$-violating domains. The glasma is a transient state of matter that connects the CGC to the QGP. As indicated above it has coherent fields, like the CGC, but it has properties similar to a plasma with strong background fields, and these fields evolve in time with the natural time scale $\tau_{s} \sim 1/Q_{s}$ which is also the proper formation time scale, $\tau^{f}_{qg}$, for quarks and gluons. Here $Q_{s}$ is the saturation scale and at RHIC $Q_{s} \sim 1$ GeV/$c$ \cite{Baier-PLB-502-51-2001}, $\tau_{s} \sim 0.2$ fm/$c$ \cite{Gyulassy-NPA-750-30-2005}.
Then through some as yet not completely understood mechanism, the glasma eventually thermalizes into a sQGP \cite{Baier-PLB-502-51-2001} with a "thermalization" time, estimated between 0.15 fm/$c$ and 1 fm/$c$ after the nuclei cross. Thus the total duration of pre-equilibrium stage is estimated as $\Delta\tau_{p-e} \simeq 1$ fm/$c$ for nucleus-nucleus collisions at RHIC energies, i.e. this is time for thermalized sQGP creation after the nuclei collision at $\tau_{0}=0$. Experimental results (see, for instance, \cite{Arsene-NPA-757-1-2005}) confirm the short duration of the pre-equilibrium stage with respect to the later thermodynamically equilibrium stages which characterized by total duration $\Delta\tau_{e} \sim 20$ fm/$c$, where the high border of the last stage correspond to the kinetic freeze-out of the hadronic gas $\tau_{fr} \sim 20$ fm/$c$. Each of these principal periods, $\tau_{0} < \tau \leq \tau_{p-e}$ and $\tau_{p-e} < \tau \leq \tau_{fr}$, contains several sub-stages and detailed description is model-dependent for evolution of strongly interacting matter during each of these two principal periods.

Jet and leading hadron measurements are believed to probe the early stages of the dense medium while soft hadronic observables deliver information on the initial state and hydrodynamic evolution of the system. But the cleanest and most sensitive probes to the pre-equilibrium phase of the nucleus-nucleus collision are photons and leptons, since they do not interact with the concurrently forming QGP medium after their production \cite{Martinez-PRC-78-034917-2008}.
The key questions discussed here are (\textit{i}) properties of sQGP and (\textit{ii}) features of QCD phase diagram.

The scientific complex with RHIC was designed and was built for investigations in QCD field specially. There were 17 successful physical runs since 2000 year. At present the only large experiment STAR continues to collect new data. The STAR setup is characterized by uniform and large acceptance, good particle identification and stable operation of the main subsystems, for instance, the main tracker -- time projection chamber (TPC) -- since first physics run in 2000 year \cite{STAR-NIM-A499-624-2003}. Table~\ref{tab-1} shows the data samples collected by STAR experiment during 17 runs at RHIC.

\begin{table}
\centering
\caption{Data samples obtained by STAR during RHIC runs.}
\label{tab-1}
\begin{tabular}{ll}
Species &
$\sqrt{s_{NN}}$, GeV\\\hline
$\mbox{p+p}^{\,\text a}$     & 22.0$^{\text b}$, 62.4, 200, 410$^{\text b}$, 500, 510 \\
$\mbox{p+Al}$              & 200 \\
$\mbox{p+Au}$              & 200 \\
$\mbox{d+Au}$              & 19.6, 39.0, 62.4, 200 \\
$\mbox{He$^{3}$+Au}$       & 200 \\
$\mbox{Al+Au}$             & 4.9$^{\text b,c}$,200 \\
$\mbox{Cu+Cu}$             & 22.4$^{\text b}$, 62.4, 200 \\
$\mbox{Cu+Au}$             & 200 \\
$\mbox{Au+Au}$             & 4.5$^{\text b,c}$,7.7, 9.2$^{\text b}$, 11.5, 14.5, 19.6, 27.0, 39.0, 54.4, 55.8$^{\text b}$, 62.4, 130, 200 \\
$\mbox{U+U}$               & 193 \\\hline
\multicolumn{2}{l}{\footnotesize $^{\text a}$ with unpolarized ($\sqrt{s}=62.4$ GeV) and with
longitudinal / transverse polarized beams;} \\
\multicolumn{2}{l}{\footnotesize $^{\text b}$ run with small integral luminosity} \\
\multicolumn{2}{l}{\footnotesize $^{\text c}$ run for STAR fixed target mode} \\
\end{tabular}
\end{table}

\section{Hard physics}\label{sec-2}
Jet quenching is considered as one of the most promising signatures of formation of sQGP and sensitive probe for transport properties of final-
state matter. Study of jet quenching with various type triggers is considered as important tool for detailed investigation of transport properties of sQGP. There are two main experimental techniques for jet studies: with help of azimuthal correlations and via full jet reconstruction.
The presence of a “trigger” particle, having $p_{\,\footnotesize{T}}$ greater than some selected value, serves as part of the selection criteria to analyze the event for a hard scattering. Within the first approach the trigger particle with high $p_{\,\footnotesize{T}}$ is selected in event, then pairs "trigger particle + associated particle" are formed and $\Delta x=x_{\,\footnotesize{t}}-x_{\,\footnotesize{a}}$ are calculated, where $x_{\,\footnotesize{t (a)}} \equiv \eta_{\,\footnotesize{t (a)}}, \phi_{\,\footnotesize{t (a)}}$ is the kinematic parameter ($\phi$ -- azimuthal angle, $\eta$ -- pseudorapidity) of trigger (associated) particle. As expect, correlations appear at $\Delta \phi \approx 0$ the peak corresponds to jet in which the both trigger and associated particles are fragmentation products of one parton and at $\Delta \phi \approx \pi$ the peak corresponds to jet in which the trigger and associated particles are fragmentation products of different (back-to-back) partons. The following quantities are calculated for particle pairs:
\begin{equation}
C_{2}(\Delta \phi)=\frac{\textstyle 1}{\textstyle N_{\,\scriptsize{\mbox{trig}}}}\int d\Delta \eta N(\Delta \phi, \Delta \eta),~~~ Y=\int d\Delta \phi C_{2}(\Delta \phi),
\label{eq:1-1}
\end{equation}
where $C_{2}(\Delta \phi)$ is the two-particle azimuthal correlation function, $Y$ -- the per-trigger yield of particle pairs. The development of experimental technique and data analysis results in the possibility for full reconstruction of hadron jets in nucleus-nucleus collisions. At present fully reconstructed jets are one of the main tools to constrain the medium transport parameters.

\subsection{Neutral probes} \label{subsec-2-1}
It was pointed out that direct photons ($\gamma_{\footnotesize{dir}}$), real or virtual, are penetrating probes for the bulk matter produced in hadronic collisions, as they do not interact strongly and they have a large mean free path \cite{Feinberg-NuovoCim-A34-391-1976}. The recent STAR physics analyses are focused on azimuthal correlations with neutral triggers ($\gamma_{\footnotesize{dir}}$ and $\pi^{0}$) produced in nucleus-nucleus collisions and on the $\gamma_{\footnotesize{dir}}$ yield.

\begin{figure}
\centering
\sidecaption
\includegraphics[width=8cm,clip]{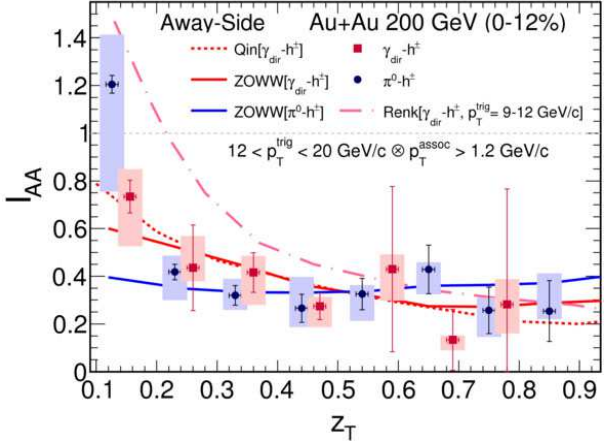}
\caption{$z_{\,\footnotesize{T}}$–dependence of $I_{\footnotesize{AA}}$ for $\pi^{0}$ and $\gamma_{\footnotesize{dir}}$ triggers (the points are slightly shifted for visibility) in most central $\mbox{Au+Au}$ collisions. The vertical lines are statistical errors, vertical extent of the boxes – systematic uncertainties. The curves represent the model predictions. Figure is taken from \cite{STAR-PLB-760-689-2016}.}
\label{fig-2-1-1}
\end{figure}

The azimuthal angular correlation of charged hadrons with respect to a direct-photon trigger was proposed as a promising probe to study the mechanisms of parton energy loss \cite{Wang-PRL-77-231-1996}. The ratio $I_{\footnotesize{AA}}(z_{\,\footnotesize{T}})=Y^{\scriptsize{\mbox{Au}}+\scriptsize{\mbox{Au}}}(z_{\,\footnotesize{T}}) / Y^{p+p}(z_{\,\footnotesize{T}})$ in $\mbox{Au+Au}$ at $\sqrt{s_{NN}}=200$ GeV has been studied with help of $\pi^{0} / \gamma_{\footnotesize{dir}} - h^{\pm}$ azimuthal correlations, where $z_{\,\footnotesize{T}}=p^{\,a}_{\,\footnotesize{T}}/p^{\,t}_{\,\footnotesize{T}}$ and it is expected $I_{\footnotesize{AA}}=1$ in the absence of medium modification. Results are shown in Fig. \ref{fig-2-1-1}. As seen $I_{\footnotesize{AA}}(z_{\,\footnotesize{T}})$ for both $\pi^{0}$ and $\gamma_{\footnotesize{dir}}$ triggers show indication on less suppression at low $z_{\,\footnotesize{T}} \in (0.1;0.2)$ than at high $z_{\,\footnotesize{T}}$ as well as similar levels of suppression for both trigger types \cite{STAR-PLB-760-689-2016}. There is no manifestation of the expected differences due to the color-factor effect and  the difference in average path lengths between $\pi^{0}$ and $\gamma_{\footnotesize{dir}}$ triggers (path-length dependence).

\begin{figure}
\centering
\sidecaption
\includegraphics[width=8cm,clip]{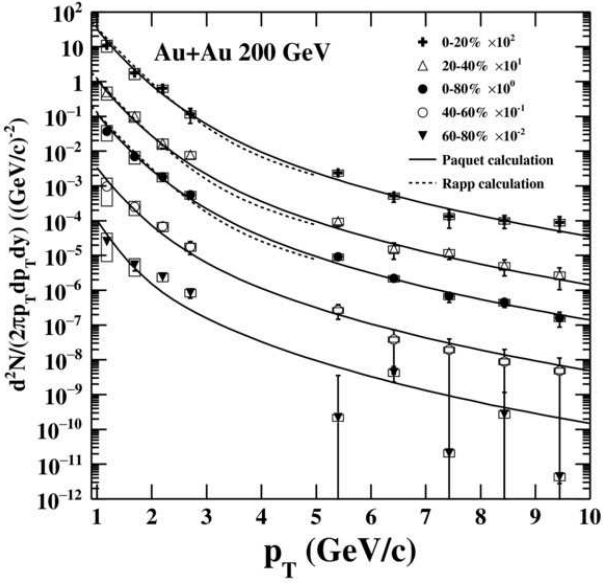}
\caption{The direct $\gamma$ invariant yield as a function of $p_{\,\footnotesize{T}}$ in comparison with model predictions. The statistical
and systematic errors are shown by bars and boxes, respectively. Figure is taken from \cite{STAR-PLB-770-451-2017}.}
\label{fig-2-1-2}
\end{figure}

Also the $\gamma_{\footnotesize{dir}}$ invariant yields are derived at midrapidity in the ranges $1 < p_{\,\footnotesize{T}} < 3$ GeV/$c$ and $5 < p_{\,\footnotesize{T}} < 10$ GeV/$c$ for $\mbox{Au+Au}$ collisions at $\sqrt{s_{NN}}=200$ GeV \cite{STAR-PLB-770-451-2017}. In Fig. \ref{fig-2-1-2} a clear excess in the $\gamma_{\footnotesize{dir}}$ invariant yield is observed in comparison with the nuclear overlap function $T_{\footnotesize{AA}}$ scaled $p+p$ reference in the first $p_{\,\footnotesize{T}}$. For $p_{\,\footnotesize{T}} > 6$ GeV/$c$ the production follows $T_{\footnotesize{AA}}$ scaling. Model calculations with contributions from thermal radiation and initial hard parton scattering are consistent within experimental uncertainties with the $\gamma_{\footnotesize{dir}}$ invariant yield in 0--20\%, 20--40\% and 40--60\% centrality bins. In 60--80\% collisions there is systematic excess data over model curve at $2 < p_{\,\footnotesize{T}} < 3$ GeV/$c$.

\subsection{Hadronic jets} \label{subsec-2-2}
\begin{figure}
\centering
\sidecaption
\includegraphics[width=8cm,clip]{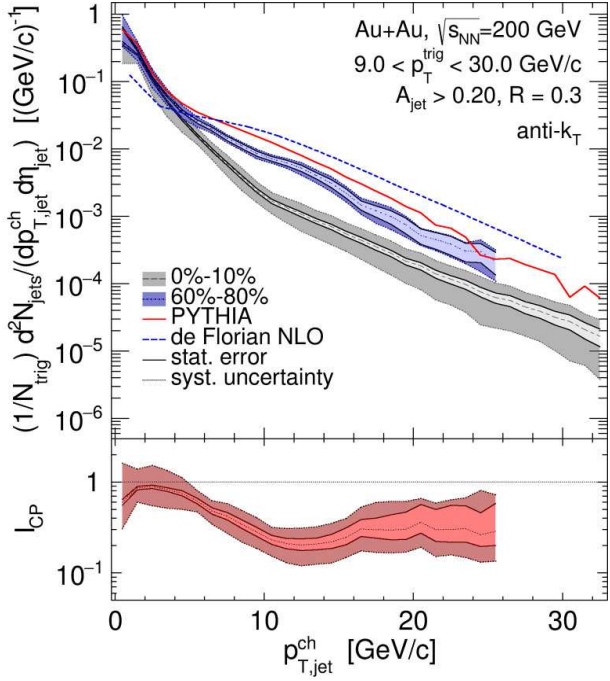}
\caption{Fully corrected distributions of $Y(p_{\,\footnotesize{T,}\,\scriptsize{\mbox{jet}}}^{\scriptsize{\mbox{ch}}})$ (upper) and its ratio $I_{\footnotesize{CP}}$ (lower) for central and peripheral $\mbox{Au+Au}$ collisions, for anti-$k_{\,\footnotesize{T}}$ jets with $R=0.3$ and area $a_{\,\scriptsize{\mbox{jet}}} > 0.2$. The upper panels also show yield for $p+p$ collisions, calculated using PYTHIA at the charged-particle level and NLO pQCD transformed to the charged-particle level. The uncertainty of the NLO calculation is not shown. Figure is taken from \cite{STAR-PRC-96-024905-2017}.}
\label{fig-2-2-1}
\end{figure}

In STAR the anti-$k_{\,\footnotesize{T}}$ algorithm with $R =0.2-0.5$ ($R^{2}=(\Delta \phi)^{2}+(\Delta \eta)^{2}$) and a low IR-cutoff ($p_{\,\footnotesize{T}} > 0.2$ GeV/$c$) is used for full jet reconstruction from charged tracks in Au+Au collisions. STAR measures the semi-inclusive distributions of reconstructed charged particle jets recoiling from a high $p_{\,\footnotesize{T}}$ trigger hadron \cite{STAR-PRC-96-024905-2017}
\begin{equation}
\frac{\textstyle 1}{\textstyle N_{\,\scriptsize{\mbox{trig}}}}\frac{\textstyle d^{2}N_{\scriptsize{\mbox{jet}}}}{\textstyle dp_{\,\footnotesize{T,}\,\scriptsize{\mbox{jet}}}^{\scriptsize{\mbox{ch}}}\,d\eta_{\scriptsize{\mbox{jet}}}}=\frac{\textstyle 1}{\textstyle \sigma^{\scriptsize{\mbox{A}}+\scriptsize{\mbox{A}} \to h+X}}\frac{\textstyle d^{2}\sigma^{\scriptsize{\mbox{A}}+\scriptsize{\mbox{A}} \to h+\scriptsize{\mbox{jet}}+X}}{\textstyle dp_{\,\footnotesize{T,}\,\scriptsize{\mbox{jet}}}^{\scriptsize{\mbox{ch}}}\,d\eta_{\scriptsize{\mbox{jet}}}},
\label{eq:1-2}
\end{equation}
where $\eta_{\scriptsize{\mbox{jet}}}$ is the pseudorapidity of  jet centroid, $p_{\,\footnotesize{T,}\,\scriptsize{\mbox{jet}}}^{\scriptsize{\mbox{ch}}}$ -- transverse momentum of jet, $\sigma^{\scriptsize{\mbox{A}}+\scriptsize{\mbox{A}} \to h+X}$ is the cross-section to generate a trigger hadron, and $d^{2}\sigma^{\scriptsize{\mbox{A}}+\scriptsize{\mbox{A}} \to h+\scriptsize{\mbox{jet}}+X} / dp_{\,\footnotesize{T,}\,\scriptsize{\mbox{jet}}}^{\scriptsize{\mbox{ch}}}\,d\eta_{J}$ is the double differential cross-section for coincidence production of a trigger hadron and a recoil jet, $\mbox{A}+\mbox{A} \equiv p+p$, $\mbox{Au+Au}$. The combinatorial background is removed by mixed-event technique. This allows the access for low-$p_{\,\footnotesize{T}}$ jets. The STAR has measured the $I_{\footnotesize{CP}}$ ratio of jet yields in central to peripheral $\mbox{Au+Au}$ events at $\sqrt{s_{NN}}=200$ GeV \cite{STAR-PRC-96-024905-2017}. Results for jet cone radius $R=0.3$ is shown in Fig. \ref{fig-2-2-1} for invariant yield (upper panel) and for $I_{\footnotesize{CP}}$ (lower panel) depends on $p_{\,\footnotesize{T,}\,\scriptsize{\mbox{jet}}}^{\scriptsize{\mbox{ch}}}$. The $I_{\footnotesize{CP}}$ shows a clear suppression of yields in central events for $p_{\,\footnotesize{T,}\,\scriptsize{\mbox{jet}}}^{\scriptsize{\mbox{ch}}} > 10$ GeV/$c$ for all cone radii $R=0.2$, 0.3, 0.4 and 0.5 under study. The horizontal shifts in the range where the $I_{\footnotesize{CP}}$ is flat ($10 < p_{\,\footnotesize{T,}\,\scriptsize{\mbox{jet}}}^{\scriptsize{\mbox{ch}}} < 20$ GeV/$c$) are $-6.3 \pm 0.6 \pm 0.8$ GeV/$c$ and $-3.8 \pm 0.5 \pm 1.8$ GeV/$c$ for $R=0.3$ and $R=0.5$, respectively. The suppression and shift, and their reduction for larger cones might indicate an out-of-cone energy transport by interaction of the jet with the medium, averaged over the recoil jet population.

\begin{figure}
\centering
\sidecaption
\includegraphics[width=8cm,clip]{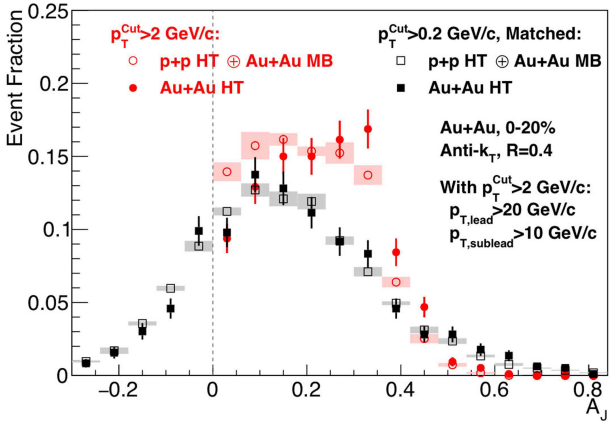}
\caption{Normalized $A_{\,\footnotesize{J}}$ distributions for $\mbox{Au+Au}$ HT data (filled symbols) and $p+p$ HT $\oplus$ $\mbox{Au+Au}$ MB (open symbols). The circles are for jets found using only constituents with $p_{\,\footnotesize{T}}^{\scriptsize{\mbox{Cut}}} > 2$ GeV/$c$ and the squares for matched jets found using constituents with $p_{\,\footnotesize{T}}^{\scriptsize{\mbox{Cut}}} > 0.2$ GeV/$c$. In all cases $R=0.4$. Statistical errors may be smaller than symbol size for $p+p$ HT $\oplus$ $\mbox{Au+Au}$ MB. Figure is taken from \cite{STAR-PRL-119-062301-2017}.}
\label{fig-2-2-2}
\end{figure}

The first dijet imbalance measurement at RHIC was made by STAR for central $\mbox{Au+Au}$ collisions at $\sqrt{s_{NN}}=200$ GeV \cite{STAR-PRL-119-062301-2017}, thus allowing a more direct comparison to jet quenching measurements at the Large Hadron Collider (LHC). The imbalance observable is defined as
\begin{equation}
A_{\,\footnotesize{J}}=\frac{\textstyle p_{\,\footnotesize{T,}\,\scriptsize{\mbox{lead}}}-p_{\,\footnotesize{T,}\,\scriptsize{\mbox{sublead}}}}{\textstyle p_{\,\footnotesize{T,}\,\scriptsize{\mbox{lead}}}+p_{\,\footnotesize{T,}\,\scriptsize{\mbox{sublead}}}}
\label{eq:1-3}
\end{equation}
where $p_{\,\footnotesize{T,}\,\scriptsize{\mbox{lead}}}$ and $p_{\,\footnotesize{T,}\,\scriptsize{\mbox{sublead}}}$ are the transverse momenta of the leading and subleading jet, respectively, in the dijets that are required to be
approximately back to back \cite{STAR-PRL-119-062301-2017}. In STAR analysis $\mbox{Au+Au}$ events at $\sqrt{s_{NN}}=200$ GeV were selected by an online high tower (HT) trigger, which required an uncorrected transverse energy of neutral hadrons $E_{\,\footnotesize{T}} > 5.4$ GeV in at least one barrel electromagnetic calorimeter (BEMC) tower. It was shown that $\mbox{Au+Au}$ HT leading jets are similar to $p+p$ HT leading jets embedded in an $\mbox{Au+Au}$ minimum bias (MB) background \cite{CMS-JHEP-0216-156-2016}. In Fig. \ref{fig-2-2-2} the $A_{\,\footnotesize{J}}$ distribution from central $\mbox{Au+Au}$ collisions for anti-$k_{\,\footnotesize{T}}$ jets with $R=0.4$ (solid points) is compared to the $p+p$ HT embedding reference ($p+p$ HT $\oplus$ $\mbox{Au+Au}$ MB, open symbols) for a jet constituents with $p_{\,\footnotesize{T}}^{\scriptsize{\mbox{Cut}}} > 2$ GeV/$c$. Dijets in central $\mbox{Au+Au}$ collisions are significantly more imbalanced than the corresponding $p+p$ dijets. The calculation in accordance with a Kolmogorov--Smirnov
test on the unbinned data \cite{Chakravarti-book-1967} supports the hypothesis that the $Au+Au$ and $p+p$ HT $\oplus$ $\mbox{Au+Au}$ MB data are not drawn from the same parent $A_{\,\footnotesize{J}}$ distributions.

\section{Soft physics} \label{sec-3}
In relativistic nuclear physics the event shape is usually studied with help of the Fourier expansion of particle distribution \cite{Voloshin-ZPC-70-665-1996}. Then with taking into account possible influence of QCD fundamental symmetries violation at finite temperature the invariant distribution of final
state particles with certain type can be written as following
\begin{equation}
E\frac{\textstyle d^{3}N}{\textstyle
d\vec{p}}=\frac{\textstyle 1}{\textstyle 2\pi} \frac{\textstyle
d^{\,2}N}{\textstyle
p_{\,\footnotesize{T}}dp_{\,\footnotesize{T}}dy}
\biggl\{1+
2\sum\limits_{n=1}^{\infty}\sum\limits_{j=1}^{2}k_{n}^{j}F_{j}\left[n\left
(\phi-\Psi_{\mbox{\scriptsize{RP}}}\right)\right] \biggr\},~~\{F\}_{j=1}^{\,2}(x) \equiv \displaystyle
\binom{\sin}{\cos}(x). \label{eq:3-1}
\end{equation}
Here $\phi$ is an azimuthal angle of particle under study,
$\Psi_{\mbox{\footnotesize{RP}}}$ -- azimuthal angle of reaction
plane, $k_{n}^{1} \equiv v_{n}=\langle \cos[n
(\phi-\Psi_{\mbox{\scriptsize{RP}}})]\rangle$ -- collective flow of $n$-th order, the
parameters $k_{n}^{2} \equiv a_{n}=\langle \sin[n
(\phi-\Psi_{\mbox{\scriptsize{RP}}})]\rangle$ describe the possible effect of $\mathcal{P/CP}$ violation in strong interactions.

\subsection{Collective flows and QCD phase diagram} \label{subsec-3-1}
Up to the present time, the main efforts are focused on the study of the low-order $v_{n}$ coefficients due to their sensitivity to the key features of the final-state matter.

The $v_{2}(\phi)/v_{2}(p)$ ratios are larger than unity at $p_{\,\footnotesize{T}} \sim 0.5$ GeV/$c$ for most central 0–30\% bin of the $\mbox{Au+Au}$ collisions at $\sqrt{s_{NN}}=200$ GeV showing an indication of breakdown of the expected mass ordering in that momentum range. This could be due to a large effect of hadronic rescattering on the proton $v_{2}$. The hypothesis has been verified by comparison with model results. The AMPT calculations show the increase of $v_{2}(\phi)/v_{2}(p)$ with growth hadronic cascade time. This is attributed to a decrease in the $v_{2}(p)$ due to an increase in hadronic rescattering, while the $\phi$-meson $v_{2}$ remains unaffected. On the other hand the ratio under discussion from UrQMD is much smaller than unity because UrQMD model lacks partonic collectivity and therefore does not fully develop the $\phi$-meson $v_{2}$ \cite{STAR-PRL-116-062301-2016}.

Similar to hadrons over the measured $p_{\,\footnotesize{T}}$ range ($0.5 \lesssim p_{\,\footnotesize{T}} < 4$ GeV/$c$), light (anti)nuclei $v_{2}(p_{\,\footnotesize{T}})$ show a monotonic rise with increasing $p_{\,\footnotesize{T}}$, mass ordering at low $p_{\,\footnotesize{T}}$, and a reduction for more central collisions. It is observed that $v_{2}$ of nuclei and antinuclei are of similar magnitude for $\mbox{Au+Au}$ collisions in domain $\sqrt{s_{NN}} \geq 39$ GeV. In general the light-nuclei $v_{2}$ follow an atomic mass number scaling indicates that the coalescence of nucleons might be the underlying mechanism of light-nuclei formation in HIC. This fact is corroborated by model calculations \cite{STAR-PRC-94-034908-2016}.

\begin{figure}
\centering
\sidecaption
\includegraphics[width=8cm,clip]{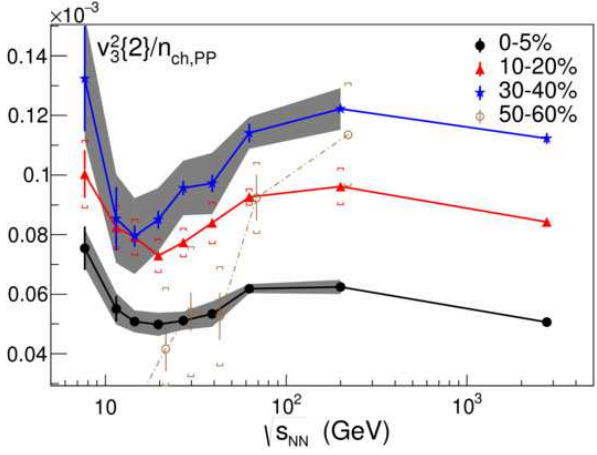}
\caption{$v_{3}^{2}\{2\}$ divided by the midrapidity, charged-particle multiplicity density per participant pair in $\mbox{Au+Au}$ and $\mbox{Pb+Pb}$ ($\sqrt{s_{NN}}=2.76$ TeV) collisions. Systematic errors are shown either as a shaded band or as thin vertical error bars with caps. Figure is taken from \cite{STAR-PRL-116-112302-2016}.}
\label{fig-3-1-1}
\end{figure}

For sufficiently central $\mbox{Au+Au}$ collisions ($N_{\scriptsize{\mbox{part}}} > 50$), the triangle flow $v_{3}$ persist down to the lowest energies studied. For more peripheral collisions, however, the correlation appears to be absent at low energies for $N_{\scriptsize{\mbox{part}}} < 50$, in agreement with AMPT model. This non-QGP model predicts a smaller $v_{3}$ value than the data, suggesting that a QGP phase may exist in more central collisions at $\sqrt{s_{NN}}=7.7$ GeV. When divided by multiplicity, $v_{3}$ shows a local minimum in the range $\sqrt{s_{NN}}=15-20$ GeV in centrality range from 0\% to 50\% (Fig. \ref{fig-3-1-1}). This feature has not been shown in any known models of HIC and could indicate an interesting trend in the pressure developed inside the system \cite{STAR-PRL-116-112302-2016}.

A finite difference in $v_{1}$ between positive and negative charged particles $(\Delta v_{1})$ was observed in the kinematic domain of $0.15 < p_{\,\footnotesize{T}} < 2$ GeV/$c$ and $|\eta| < 1$. The $\Delta v_{1}$ seems to increase with $p_{\,\footnotesize{T}}$. The $v_{1}$ results from $\mbox{Au+Au}$ collisions show much smaller values compared to those in $\mbox{Cu+Au}$. These results are consistent with the presumption of a strong, initial electric field in asymmetric collisions. The $p_{\,\footnotesize{T}}$ dependence of $\Delta v_{1}$ is qualitatively described by the parton--hadron--string--dynamics (PHSD) model at $p_{\,\footnotesize{T}} < 2$ GeV/$c$. However, the magnitude of $\Delta v_{1}$ is smaller by a factor of 10 than the model predictions \cite{STAR-PRL-118-012301-2017}. This may indicate that most of the quarks and antiquarks have not yet been created within the lifetime of the electric field ($t \leq 0.25$ fm/$c$).

\begin{figure}
\centering
\sidecaption
\includegraphics[width=8cm,clip]{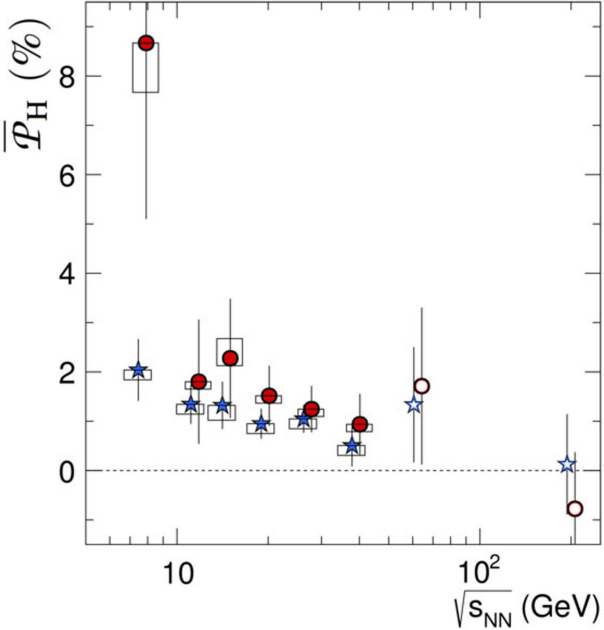}
\caption{The average polarization for $\mbox{H} \equiv \Lambda (\bar{\Lambda})$ in $\mbox{Au+Au}$ collisions. The results for $\Lambda$ (stars) and $\bar{\Lambda}$ (circles) from 20-–50\% central collisions are plotted as a function of $\sqrt{s_{NN}}$. Error bars represent statistical uncertainties only, while boxes represent systematic uncertainties. The results of the present study ($\sqrt{s_{NN}} < 40$ GeV), indicated by
filled symbols, are shown together with those reported earlier \cite{STAR-PRC-76-024915-2007} for 62.4 GeV and 200 GeV collisions, indicated by open symbols and for which only statistical errors are plotted. Figure is taken from \cite{STAR-Nature-548-62-2017}.}
\label{fig-3-2-1}
\end{figure}

\subsection{Hyperon polarization and rotating sQGP} \label{subsec-3-2}
Non-central nuclear collisions have angular momentum on the order of $10^{3}\hbar$, and shear forces generated by the interpenetrating nuclei may generate a clear vortical structure. The $\Lambda$ and $\bar{\Lambda}$ are used to measure the global hyperon polarization in non-central $\mbox{Au+Au}$ collisions because they are “self analyzing”, i.e., for instance, in the decay $\Lambda \to p+\pi^{-}$ the $p$ tends to be emitted along the spin direction of the parent $\Lambda$. At present due to the limited statistics collected by STAR, only the average projection of polarization on overall angular momentum was extracted. In Fig. \ref{fig-3-2-1} the average polarization for $\Lambda$ ($\bar{\Lambda}$) hyperons is shown as a function of $\sqrt{s_{NN}}$ for semi-central $\mbox{Au+Au}$ collisions \cite{STAR-Nature-548-62-2017}. At $\sqrt{s_{NN}} < 200$ GeV, a positive polarization is observed for $\Lambda$ ($\bar{\Lambda}$) hyperons. The data are statistically consistent with the hypothesis of energy-independent polarizations of 1.08 $\pm$ 0.15 (stat) $\pm$ 0.11 (sys) and 1.38 $\pm$ 0.30 (stat) $\pm$ 0.13 (sys) per cent for $\Lambda$ and $\bar{\Lambda}$, respectively \cite{STAR-Nature-548-62-2017}.
The $\sqrt{s_{NN}}$-averaged polarizations indicate a vorticity magnitude of $\omega=(9 \pm 1) \times 10^{21}$ s$^{-1}$, which is estimated using the hydrodynamic relations and takes into account the “feed-down” contributions. This extremely large estimation of $\omega$ evidences the sQGP is characterized by complex and developed vortical structure.

\section{Heavy quarks in hot environment}\label{sec-4}
Heavy quark masses are mostly uncharged by chiral symmetry restoration. Heavy flavor (HF) particles are produced early in collisions due to hard scattering processes. Therefore they allow the study of whole space-time evolution of sQGP. Study of heavy quark diffusion is essential for better understanding of transport properties of sQGP and underlying fluid of light partons. Precise measurements of $c$- and $b$-quarks energy loss are crucial for understanding of parton interactions with the hot environment.

\subsection{Bottomonia in sQGP} \label{subsec-4-1}
\begin{figure}
\centering
\includegraphics[width=14cm,clip]{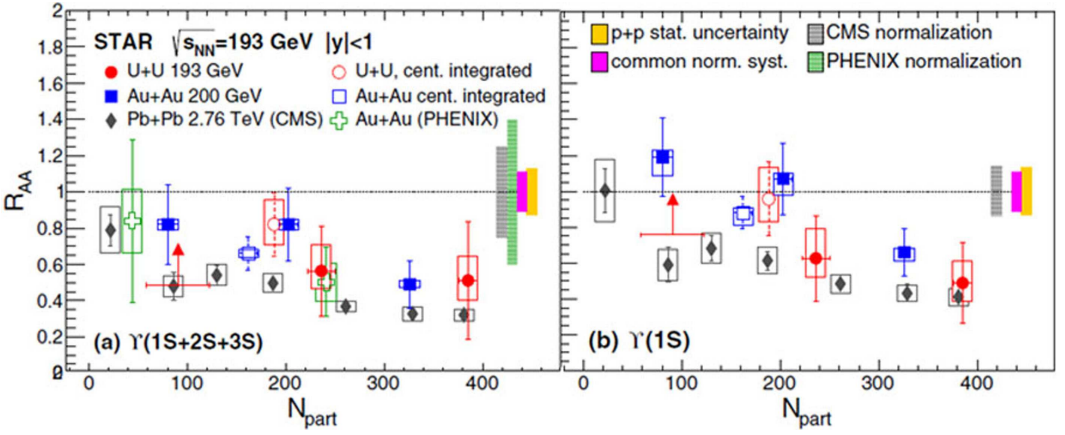}
\caption{$\Upsilon$(1S+2S+3S) (a) and $\Upsilon$(1S) (b) nuclear modifi-
cation factor $R_{\footnotesize{AA}}$ as a function of $N_{\footnotesize{part}}$ in relativistic heavy ion collisions. The data points in the 30—60\% centrality bin have large statistical and systematical uncertainties, providing little constraint on $R_{\footnotesize{AA}}$. Therefore the 95\% lower confidence bound is only indicated for the 30—60\% centrality $\mbox{U+U}$ data.
Data are from \cite{STAR-PRC-94-064904-2016}.}
\label{fig-4-1-1}
\end{figure}

The novel results have been obtained by STAR for production of $\Upsilon$ states in heavy ion collisions. The nuclear modification factor $R_{\footnotesize{AA}}$ depends on number of participants $N_{\footnotesize{part}}$ is shown in Fig. \ref{fig-4-1-1} for various $\Upsilon$ states, where
\begin{equation}
R_{\footnotesize{AA}}%=\frac{\textstyle 1}{\textstyle \langle N_{\footnotesize{bin}}\rangle}\frac{\textstyle d^{2}N^{\scriptsize{\mbox{A}}+\scriptsize{\mbox{A}}}/dp_{\,\footnotesize{T}}d\eta}{\textstyle d^{2}N^{p+p}/dp_{\,\footnotesize{T}}d\eta}
=\frac{\textstyle 1}{\textstyle T_{\footnotesize{AA}}}\frac{\textstyle d^{2}N^{\scriptsize{\mbox{A}}+\scriptsize{\mbox{A}}}/dp_{\,\footnotesize{T}}d\eta}{\textstyle d^{2}\sigma^{p+p}/dp_{\,\footnotesize{T}}d\eta}.
\label{eq:4-1}
\end{equation}

\begin{figure}
\centering
\includegraphics[width=14cm,clip]{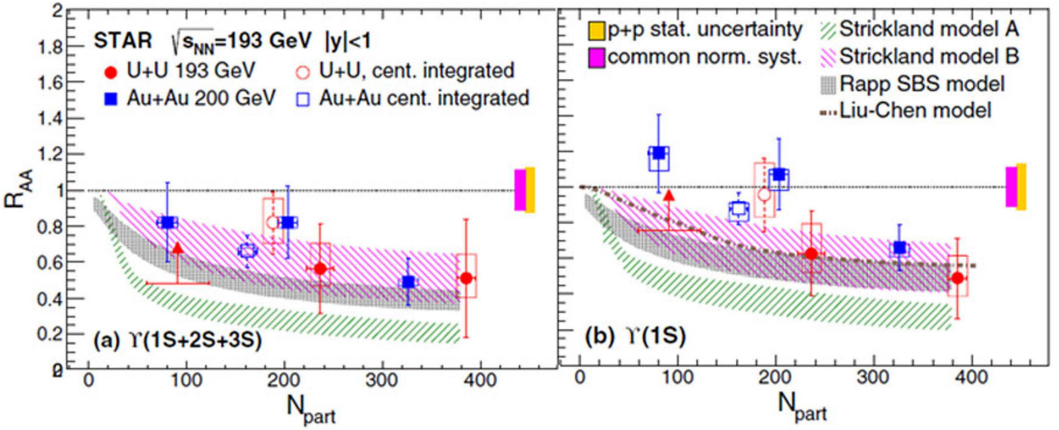}
\caption{$\Upsilon$(1S+2S+3S) (a) and $\Upsilon$(1S) (b) $R_{\footnotesize{AA}}$ depends on $N_{\footnotesize{part}}$ in
$\sqrt{s_{NN}}=193$ GeV $\mbox{U+U}$ collisions, compared to different models. The 95\% lower confidence bound is indicated for the 30-–60\% centrality $\mbox{U+U}$ data. Data and model results are taken from \cite{STAR-PRC-94-064904-2016}.}
\label{fig-4-1-2}
\end{figure}

The trend marked by the points $ R_{\footnotesize{AA}}(N_{\footnotesize{part}})$ for $\mbox{Au+Au}$ is confirmed by the $\mbox{U+U}$ data. There is strong suppression of the $\Upsilon$(2S+3S) states in $\mbox{U+U}$ as well as in $\mbox{Au+Au}$ collisions at highest available RHIC energies. There is neither a significant difference between the results in any of the centrality bins,
nor any evidence of a sudden increase in suppression in central $\mbox{U+U}$ compared to the central $\mbox{Au+Au}$ data. But one can note the precision of the current measurement does not exclude a moderate drop in $R_{\footnotesize{AA}}$. These STAR data have been compared with several models for upsilon behavior in hot environment (Fig. \ref{fig-4-1-2}). As seen the Rapp model included CNM effects, dissociation of bottomonia in the hot ($T = 330$ MeV) medium and regeneration for strongly bound scenario (SBS) for the internal-energy-based heavy quark potential describe the STAR data within uncertainties as well as other internal-energy-based models with an initial central temperature $428 < T < 442$ MeV (noted as “model B” in Fig. \ref{fig-2-1-2}) and $T = 340$ MeV (“Liu-Chen model”). The “model A” with the free-energy-based potential corresponding to a more weakly bound scenario (WBS) tends to underpredict the $R_{\footnotesize{AA}}$ especially for the $\Upsilon$(1S).

\subsection{Collectivity for heavy flavor hadrons} \label{subsec-4-2}
The heavy flavor tracker (HFT) is the newest member of the STAR detector subsystems. It is made using thinned monolithic active pixel sensor technology. STAR HFT includes pixel detector (PXL), intermediate silicon tracker (IST) and silicon strip detector (SSD). The HFT is characterized by low material budget and large acceptance coverage ($|\eta| < 1$, $0 < \phi < 2\pi$). The HFT makes it possible for the first time to directly track the decay products of hadrons comprised of charm and bottom quarks with high pointing resolution ($\sim 30$ $\mu$m for particles with $p \geq 1.5$ GeV/$c$).

\begin{figure}
\centering
\includegraphics[width=14cm,clip]{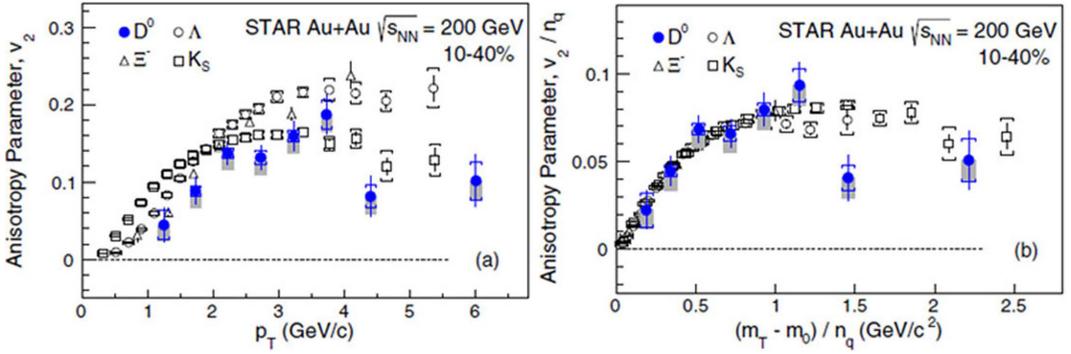}
\caption{(a) $v_{2}$ as a function of $p_{\,\footnotesize{T}}$ and (b) $v_{2}/n_{q}$ dependence on $(m_{\,\footnotesize{T}}-m_{0})/n_{q}$ for $D^{0}$ compared with earlier measurements for strange particles. The vertical bars and brackets represent statistical and systematic uncertainties, respectively, and the gray bands represent the estimated nonflow contribution. Data are from \cite{STAR-PRL-118-212301-2017}.}
\label{fig-4-2-1}
\end{figure}

\begin{figure}
\centering
\sidecaption
\includegraphics[width=8cm,clip]{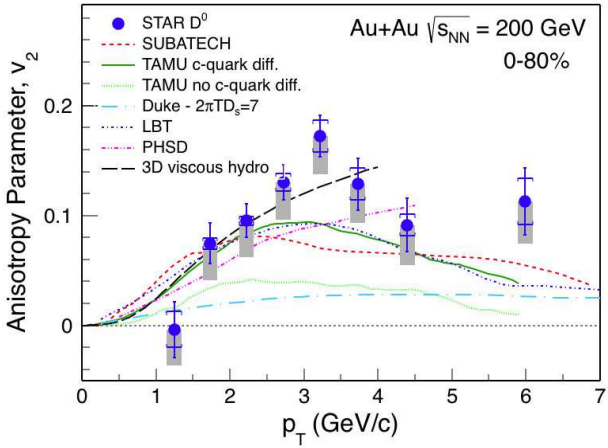}
\caption{$v_{2}$ as a function of $p_{\,\footnotesize{T}}$ for $D^{0}$ in minimum bias $\mbox{Au+Au}$ collisions compared with model calculations. Figure is taken from \cite{STAR-PRL-118-212301-2017}.}
\label{fig-4-2-2}
\end{figure}

The clear mass ordering is obtained for elliptic flow in domain $p_{\,\footnotesize{T}} < 2$ GeV/$c$ for various hadrons including $D^{0}$ mesons (Fig. \ref{fig-4-2-1}). For larger $p_{\,\footnotesize{T}} > 2$ GeV/$c$, the $D^{0}$ meson $v_{2}$ follows that of other light mesons indicating a significant $c$-quark flow at highest RHIC energy. The elliptic flow for $D^{0}$ falls into the same universal trend as all other light hadrons, in particular, for the range $(m_{\,\footnotesize{T}}-m_{0})/n_{q} < 1$ GeV/$c^{2}$. This suggests that $c$-quarks have gained significant flow through interactions with the sQGP in the collisions under consideration. The comparison between STAR results and phenomenological calculations is shown in Fig. \ref{fig-4-2-2}. The TAMU model describes the data only with no $c$-quark diffusion. A 3D viscous hydrodynamic simulation with $\eta/s=0.12$ tuned to describe $v_{2}$ for light hadrons predicts elliptic flow values for $D^{0}$ that is consistent with data for $p_{\,\footnotesize{T}} < 4$ GeV/$c$. This suggests that $c$-quarks have achieved thermal equilibrium in these collisions. The statistical significance test was performed for the consistency between data and each model quantified by $\chi^{2}/\mbox{ndf}$. The 3D viscous hydro models shows the best $\chi^{2}/\mbox{ndf}=0.73$ for the measured $p_{\,\footnotesize{T}}$ region. The models that can describe both the $R_{\footnotesize{AA}}$ and $v_{2}$ data include the temperature-dependent charm diffusion coefficient $\kappa=2\pi TD_{s}$ in the range of $\sim 2-12$. The $\kappa$ predicted by lattice QCD calculations fall in the same range \cite{STAR-PRL-118-212301-2017}.

\section{Future plans}\label{sec-5}
The STAR Collaboration works intensively under development the physics program and detector upgrades for nearest RHIC runs (2018--2020) as well as for future studies of strong interactions after 2020 year. Especially, STAR’s highest scientific priority for run 18 is the successful realization of the isobaric collision program ($\mbox{Rb+Rb}$, $\mbox{Zr+Zr}$ at $\sqrt{s_{NN}}=200$ GeV) for clarification the interpretation of measurements related to chiral effects, in particular chiral magnetic effect (CME).
%Collisions of isobaric nuclei ($Z_{\scriptsize{\mbox{Rb}}}=44$, %$Z_{\scriptsize{\mbox{Zr}}}=40$) present a unique opportunity to vary the initial %magnetic field ($B \propto Z^{2}$) and then separate the CME-signal and the %flow-related background.

\subsection{Nearest upgrades for stage II of the BES program at RHIC} \label{subsec-5-1}
As expected there will important improvement of RHIC for stage II of the BES. Low energy electron cooling at RHIC allows the increase of luminosity by factors 3--10 for BES energy ranges in both the collider mode ($\sqrt{s_{NN}}=7.7, 11.5, 14.5$ and 19.6 GeV) and the fixed target mode ($\sqrt{s_{NN}}=3.0, 3.5$ and 4.5 GeV) for heavy ($\mbox{Au+Au}$) ion collisions. Upgrade for STAR promises several improvements for physics analyses like significant decrease of statistical and systematics
uncertainties, advanced particle identification (PID) capability, broader kinematic coverage etc. The three detector subsystems will be made for stage II of BES at RHIC: upgrade of inner sectors of TPC (iTPC), event plane detector (EPD) and endcap for time-of-light (eTOF).

\begin{figure}
\centering
\sidecaption
\includegraphics[width=8cm,clip]{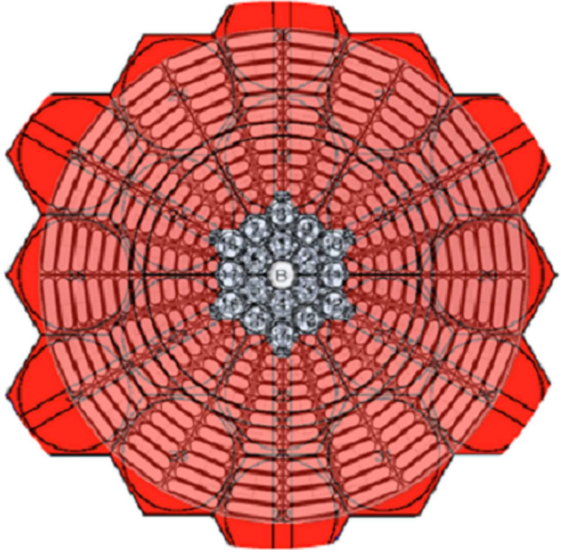}
\caption{Schematic view of EPD with respect to the existing beam-beam counter (BBC) subsystem (central part). There will be 372 channels for EPD instead of 16 present channels for BBC.}\label{fig-5-1-1}
\end{figure}

Replace all inner sector of TPC will improve the continues coverage (40 pad rows instead of 13 at present) and the dE/dx on 15–30\%. Also coverage on $|\eta|$ will be extended from 1.0 to 1.5. As planed, iTPC will be characterized by better momentum resolution and lowers $p_{\,\footnotesize{T}}$ cut from 125 MeV/c to 60 MeV/c. Beam-beam counter (BBC) will be replaced by new EPD with significantly higher azimuthal and radial segmentation (Fig. \ref{fig-5-1-1}). The EPD is crucially important for study of $\Lambda$ ($\bar{\Lambda}$) polarization in nuclear collisions as well as for BES physics due to independent event plane measurement. As expected, the EPD will provides better trigger and background reduction, extends $|\eta|$ coverage from $3.3 < |\eta| < 5.0$ for present BBC to $2.1 < |\eta| < 5.1$. The EPD will allows the measurement the centrality and event plane at forward rapidity and improvement for event plane resolution. During the Run 17 the first 1/8 part of EPD was installed and tested successfully. The eTOF is the joint project between STAR and CBM. This detector is important for fixed target program and complements the existing barrel TOF. The major improvements for eTOF installation are following: midrapidity particle identification in fixed target mode, extension of identification capabilities for $\pi$, $p$ and $K$, additional for barrel TOF ($|\eta| < 0.9$) coverage: $-1.5 < \eta < -1.1$.

\subsection{Towards cold QCD: STAR after 2020} \label{subsec-5-2}
\begin{figure}
\centering
\sidecaption
\includegraphics[width=9.5cm,clip]{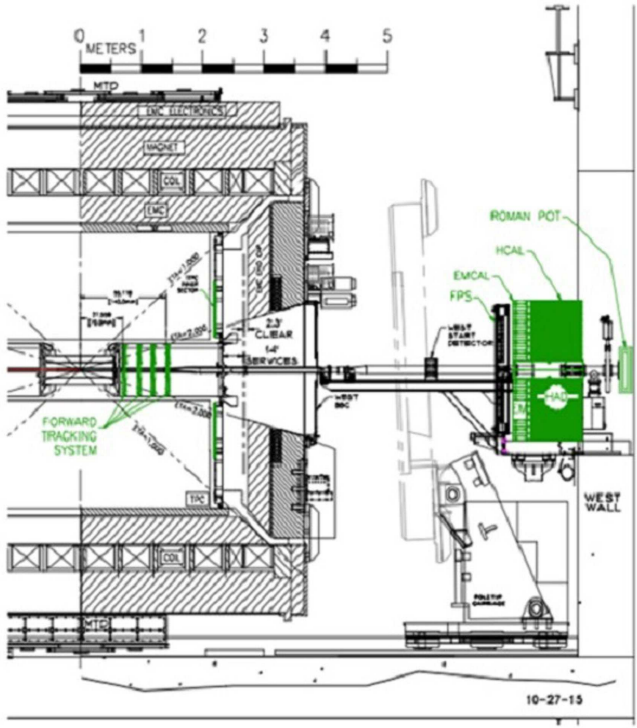}
\caption{Scheme for location of the forward calorimeter system (FCS) at the west side of the STAR detector. Figure is taken from \cite{STAR-proposal-FCS-FTS-2017}.}
\label{fig-5-2-1}
\end{figure}

The sQGP created in the final state of high energy nucleus-nucleus collision can be considered as the incoherent thermal limit of QCD matter at high temperatures while the CGC in the initial state is the coherent limit of QCD at high energies. Since the sQGP has to be created at modern colliders from the interaction of initial nuclear enhanced coherent chromo electric magnetic fields through glasma, both limiting forms of QCD matter as well as transient glasma state should be studied in detail. The forward heavy ion collisions provide the unique opportunity to study the various saturation regimes. The RHIC cold QCD plan requires an upgrade to the forward rapidity ($2.5 < \eta < 4.5$) detection capabilities of STAR. According to the \cite{STAR-proposal-FCS-FTS-2017} it is suggested the development of the STAR calorimeter and tracking subsystems for forward region (Fig. \ref{fig-5-2-1}). This upgrade allows the exploration of cold QCD physics in very high and low regions of Bjorken $x_{\,\footnotesize{B}}$ as well as the longitudinal structure of the initial state and the temperature dependent transport properties of matter in nuclear collisions with help of measurements of azimuthal asymmetries, dihadron and $h/\gamma$-jet correlations, asymmetries in production of hadrons and jets, $R_{\,\footnotesize{pA}}$ for $\gamma_{\footnotesize{dir}}$ and Drell–Yan process.

\section{Summary}\label{sec-6}
During the last years many important results have been obtained within heavy ion program of the STAR experiment. Important progress was achieved for study of hard processes in nucleus-nucleus collisions, namely, azimuthal correlations with neutral triggers show the absence of path-length dependence for neutral pions and direct photons; semi-inclusive jet production demonstrate the possible reduction of medium-induced energy transport to large angles at RHIC with respect to the LHC.
The STAR results for dijet imbalance in $\mbox{Au+Au}$ collisions at $\sqrt{s_{NN}}=200$ GeV are the first indication that at RHIC energies it is possible to select a sample of reconstructed dijets that clearly lost energy via interactions with the medium but whose lost energy reemerges as soft constituents accompanied with a small, but significant, broadening of the jet structure compared to $p+p$ fragmentation. The above observations are consistent with the qualitative expectations of perturbative QCD like radiative energy loss in the hot, dense medium
created at RHIC. In soft physics field it was found $\Lambda$($\bar{\Lambda}$) hyperons are polarized in nuclear collisions. The estimation of $\omega$ can be considered as evidence that quark-gluon matter produced at RHIC is most vortical fluid among known. Directed flow indicates on the presence of a strong, initial electric field in asymmetric nuclear collisions. Light (anti)nuclei and $D^{0}$ mesons show the collective behavior which agrees with general trends. Quarkonia measurements in heavy ion collisions consistent with the expectations from the sequential melting hypothesis.

Thus the present stage in studies of nuclear collisions in STAR experiment is the transition from the qualitative statements to the quantitative understanding in relativistic nuclear physics.

There are detailed plans for both the nearest runs (18 and 19) at RHIC and the future investigations after 2020 year. For the first case detector upgrades are in progress for precise study of sQGP within stage II of the BES. The future modification of the STAR setup after 2020 aims on the study of effects of QCD cold matter, longitudinal structure of initial conditions.
%
% BibTeX or Biber users please use (the style is already called in the class, ensure that the "woc.bst" style is in your local directory)
% \bibliography{name or your bibliography database}

\begin{thebibliography}{}
%1
\bibitem{Ludlam-PT-56-48-2003}
T. Ludlam and L. McLerran, Phys. Today \textbf{56}, 48 (2003).
%2
\bibitem{Kovner-PRD-52-6231-1995}
A. Kovner, L.D. McLerran, H. Weigert, Phys. Rev. D\textbf{52}, 6231 (1995); 3809 (1995); A. Krasnitz, R. Venugopalan, Phys. Rev. Lett \textbf{84}, 4309 (2000); A. Krasnitz, R. Venugopalan, Nucl. Phys. B\textbf{557}, 237 (1999); A. Krasnitz, Y. Nara, R. Venugopalan, Phys. Rev. Lett. \textbf{87}, 192302 (2001).
%3
\bibitem{Iancu-arXiv-hep-ph-0303204-2003}
E. Iancu, R. Venugopalan, arXiv: hep-ph/0303204 (2003).
%4
\bibitem{Krasnitz-PRL-86-1717-2001}
A. Krasnitz, R. Venugopalan, Phys. Rev. Lett. \textbf{86}, 1717 (2001); T. Lappi, Phys. Rev. C\textbf{67}, 054903 (2003); T. Lappi, L. McLerran, Nucl. Phys. A\textbf{772}, 200 (2006).
%5
\bibitem{McLerran-arXiv-0911.2987}
L. McLerran, arXiv: 0911.2987 [hep-ph] (2009).
%6
\bibitem{Baier-PLB-502-51-2001}
R. Baier et al., Phys. Lett. B\textbf{502}, 51 (2001); \textit{ibid} \textbf{539}, 46 (2002).
%7
\bibitem{Gyulassy-NPA-750-30-2005}
M. Gyulassy, L. McLerran, Nucl. Phys. A\textbf{750}, 30 (2005).
%8
\bibitem{Arsene-NPA-757-1-2005}
I. Arsene et al. (BRAHMS Collaboration), Nucl. Phys. A\textbf{757}, 1 (2005); K. Adcox et al. (PHENIX Collaboration), 184; B.B. Back et al. (PHOBOS Collaboration), 28; J. Adams et al. (STAR Collaboration), 102.
%9
\bibitem{Martinez-PRC-78-034917-2008}
M. Martinez, M. Strickland, Phys. Rev. C\textbf{78}, 034917 (2008); Eur. Phys. J. C\textbf{61}, 905 (2009).
%10
\bibitem{STAR-NIM-A499-624-2003}
K. H. Ackermann et al. (STAR Collaboration), Nucl. Inst. \& Meth. A\textbf{499}, 624 (2003).
%11
\bibitem{Feinberg-NuovoCim-A34-391-1976}
E. L. Feinberg, Nuovo Cim. A\textbf{34}, 391 (1976).
%12
\bibitem{Wang-PRL-77-231-1996}
X.-N. Wang, Z. Huang, and I. Sarcevic, Phys. Rev. Lett. \textbf{77}, 231 (1996).
%13
\bibitem{STAR-PLB-760-689-2016}
L. Adamczyk et al. (STAR Collaboration), Phys. Lett. B\textbf{760}, 689 (2016).
%14
\bibitem{STAR-PLB-770-451-2017}
L. Adamczyk et al. (STAR Collaboration), Phys. Lett. B\textbf{770}, 451 (2017).
%15
\bibitem{STAR-PRC-96-024905-2017}
L. Adamczyk et al. (STAR Collaboration), Phys. Rev. C\textbf{96}, 024905 (2017).
%16
\bibitem{STAR-PRL-119-062301-2017}
L. Adamczyk et al. (STAR Collaboration), Phys. Rev. Lett. \textbf{119}, 062301 (2017).
%17
\bibitem{Voloshin-ZPC-70-665-1996}
S. A. Voloshin and Y. Zhang, Z. Phys. C\textbf{70}, 665 (1996).
%18
\bibitem{STAR-PRL-116-062301-2016}
L. Adamczyk et al. (STAR Collaboration), Phys. Rev. Lett. \textbf{116}, 062301 (2016).
%19
\bibitem{STAR-PRC-94-034908-2016}
L. Adamczyk et al. (STAR Collaboration), Phys. Rev. C\textbf{94}, 034908 (2016).
%20
\bibitem{STAR-PRL-116-112302-2016}
L. Adamczyk et al. (STAR Collaboration), Phys. Rev. Lett. \textbf{116}, 112302 (2016).
%21
\bibitem{STAR-PRL-118-012301-2017}
L. Adamczyk et al. (STAR Collaboration), Phys. Rev. Lett. \textbf{118}, 012301 (2017).
%22
\bibitem{STAR-Nature-548-62-2017}
L. Adamczyk et al. (STAR Collaboration), Nature \textbf{548}, 62 (2017).
%23
\bibitem{STAR-PRC-76-024915-2007}
B. I. Abelev et al. (STAR Collaboration), Phys. Rev. C\textbf{76}, 024915 (2007).
%24
\bibitem{CMS-JHEP-0216-156-2016}
V. Khachatryan et al. (CMS Collaboration), J. High Energy Phys. \textbf{02}, 156 (2016).
%25
\bibitem{Chakravarti-book-1967}
I. M. Chakravarti, R. G. Laha, and J. Roy, \textit{Handbook
of Methods of Applied Statistics} (John Wiley \& Sons, New York, 1967), V. \textbf{1}, pp. 392–394.
%26
\bibitem{STAR-PRC-94-064904-2016}
L. Adamczyk et al. (STAR Collaboration), Phys. Rev. C\textbf{94}, 064904 (2016).
%27
\bibitem{STAR-PRL-118-212301-2017}
L. Adamczyk et al. (STAR Collaboration), Phys. Rev. Lett. \textbf{118}, 212301 (2017).
%28
\bibitem{STAR-proposal-FCS-FTS-2017}
\textit{The STAR forward calorimeter system and forward tracking system}. Proposal, May 2017.

\end{thebibliography}
%
% Non-BibTeX users please use
%

\end{document}